\documentclass[12pt,preprint]{aastex}
\shorttitle{Discovery of Four Lensed  Double Quasars}
\shortauthors{INADA ET AL.}
\begin{document}
%%%%%%%%%%%%%%%%%%%%%%%%%%%%%%%%%%%%%%%%%%%%%%%%%%%%%%%%%%%%%%%%%%%%%%
%%%%%%%%%%%%%%%%%%%%%%%%%%%%%%%%%%%%%%%%%%%%%%%%%%%%%%%%%%%%%%%%%%%%%%
\title{Discovery of Four Doubly Imaged Quasar Lenses from the Sloan Digital Sky Survey}  
%%%%%%%%%%%%%%%%%%%%%%%%%%%%%%%%%%%%%%%%%%%%%%%%%%%%%%%%%%%%%%%%%%%%%%
%%%%%%%%%%%%%%%%%%%%%%%%%%%%%%%%%%%%%%%%%%%%%%%%%%%%%%%%%%%%%%%%%%%%%%
%
%%%%%%%%%%%%%%%%%%%%%%%%%%%%%%%%%%%%%%%%%%%%%%%%%%%%%%%%%%%%%%%%%%%%%%
\author{
Naohisa Inada,\altaffilmark{1}
Masamune Oguri,\altaffilmark{2,3}
Cristian E. Rusu,\altaffilmark{4,5}
Issha Kayo,\altaffilmark{6} 
and 
Tomoki Morokuma\altaffilmark{7}
}

\altaffiltext{1}{Department of Physics, Nara National College of Technology, 
                 Yamatokohriyama, Nara 639-1080, Japan.}                  
\altaffiltext{2}{Department of Physics, Graduate School of Science, University 
                 of Tokyo, Hongo 7-3-1, Bunkyo-ku, Tokyo 113-0033, Japan.}                  
\altaffiltext{3}{Kavli Institute for the Physics and Mathematics of the Universe, 
                 The University of Tokyo, 5-1-5 Kashiwa-no-ha, Kashiwa, 
                 Chiba 277-8568, Japan.}                  
\altaffiltext{4}{Optical and Infrared Astronomy Division, National Astronomical 
                 Observatory of Japan, 2-21-1, Osawa, Mitaka, Tokyo 181-8588, 
                 Japan.}                  
\altaffiltext{5}{Department of Astronomy, Graduate School of Science, University 
                 of Tokyo 7-3-1, Hongo Bunkyo-ku, Tokyo 113-0033, Japan.}                  
\altaffiltext{6}{Department of Physics, Toho University, Funabashi, Chiba 274-8510, 
                 Japan.}         
\altaffiltext{7}{Institute of Astronomy, Faculty of Science, University
                 of Tokyo, 2-21-1 Osawa, Mitaka, Tokyo 181-0015, Japan.}          

\begin{abstract}
We report the discovery of four doubly imaged quasar lenses. All the four
systems are selected as lensed quasar candidates from the Sloan
Digital Sky Survey data. We confirm their lensing hypothesis with
additional imaging and spectroscopic follow-up observations. The discovered
lenses are SDSS~J0743+2457 with the source redshift $z_s=2.165$, the
lens redshift $z_l=0.381$, and the image separation
$\theta=1\farcs034$, SDSS~J1128+2402 with $z_s=1.608$ and
$\theta=0\farcs844$, SDSS~J1405+0959 with $z_s=1.810$, $z_l\approx
0.66$, and $\theta=1\farcs978$, and SDSS~J1515+1511 with $z_s=2.054$,
$z_l=0.742$, and $\theta=1\farcs989$. It is difficult to estimate 
the lens redshift of SDSS~J1128+2402 from the current data. 
Two of the four systems
(SDSS~J1405+0959 and SDSS~J1515+1511) are included in our final
statistical lens sample to derive constraints on dark energy and the
evolution of massive galaxies.
\end{abstract}


%%%%%%%%%%%%%%%%%%%%%%%%%%%%%%%%%%%%%%%%%%%%%%%%
%%%%%%%%%%%%%%%%%%%%%%%%%%%%%%%%%%%%%%%%%%%%%%%%
%%%%%%%%%%%%%%%%%%%%%%%%%%%%%%%%%%%%%%%%%%%%%%%%
\section{Introduction}\label{sec:intro}
%%%%%%%%%%%%%%%%%%%%%%%%%%%%%%%%%%%%%%%%%%%%%%%%
%%%%%%%%%%%%%%%%%%%%%%%%%%%%%%%%%%%%%%%%%%%%%%%%
%%%%%%%%%%%%%%%%%%%%%%%%%%%%%%%%%%%%%%%%%%%%%%%%

Gravitationally lensed quasars have proven to be a useful tool for a
variety of astrophysical and cosmological studies, as described e.g., 
in \citet{kochanek06}. Applications of quasar lenses include
constraints  on cosmological parameters
\citep[e.g.,][]{refsdal64,turner90,fukugita90} and measurements of
mass distributions for galaxy- and cluster-scale halos
\citep[e.g.,][]{kochanek91,mao98,williams04,oguri13}, interstellar
medium associated with lensing or intervening galaxies
\citep[e.g.,][]{curran07,cooke10}, and the structure of quasar
accretion disk and outflow \citep[e.g.,][]{poindexter08,misawa13}.  
Many of these applications take advantage of the time-variable nature of
quasars and therefore are not available for galaxy-galaxy strong
lenses. The use of these quasar lenses can be enhanced by constructing
a homogeneous lens sample with a well-characterized selection function.
For instance, the Cosmic Lens All Sky Survey
\citep[CLASS;][]{myers03,browne03} has provided a complete sample of
22 strong lenses selected among $\sim$16,000 radio sources, which has
been used for a number of cosmological and astrophysical studies
\citep[e.g.,][]{chae06}.   

We have completed the SDSS Quasar Lens Search 
\citep[SQLS;][]{oguri06,oguri08,oguri12,inada08,inada10,inada12},
which is the current largest lensed quasar survey using optical data
of the Sloan Digital Sky Survey \citep[SDSS;][]{york00}. As a result, we have
discovered more than 40 lensed quasars including four lensed quasars
presented in this paper. We have also rediscovered many previously known
lensed quasars located in the SDSS footprint \citep{walsh79,weymann80,
surdej87,magain88,bade97,oscoz97,schechter98,myers99,morgan01,winn02,reimers02, 
jackson08,jackson09,jackson12}. The total number of lensed quasars
identified in the SDSS data is 62\footnote{The gravitational lens nature 
of an object is not secure \citep{rusu13}.} \citep{inada12}. The statistical
analysis for SQLS lenses to constrain both cosmological parameters and
the evolution of massive galaxies has been reported in \citet{oguri12}. 

In this paper, we report the discovery of four lensed quasars from the
SQLS. Since the description of the four lensed quasars in \citet{inada12} is 
so brief, we show the detailed properties of the four systems here. 
We briefly describe the SDSS data and our algorithm to locate
lensed quasar candidates in \S~\ref{sec:sdss}. We then present
follow-up observations to confirm their lensing hypotheses in
\S~\ref{sec:follow-up}. We carry out mass modeling in \S~\ref{sec:model}. 
Our result is summarized in \S~\ref{sec:summary}. 
 
%%%%%%%%%%%%%%%%%%%%%%%%%%%%%%%%%%%%%%%%%%%%%%%%
%%%%%%%%%%%%%%%%%%%%%%%%%%%%%%%%%%%%%%%%%%%%%%%%
%%%%%%%%%%%%%%%%%%%%%%%%%%%%%%%%%%%%%%%%%%%%%%%%
\section{Candidate Selection from the Sloan Digital Sky Survey}\label{sec:sdss} 
%%%%%%%%%%%%%%%%%%%%%%%%%%%%%%%%%%%%%%%%%%%%%%%%
%%%%%%%%%%%%%%%%%%%%%%%%%%%%%%%%%%%%%%%%%%%%%%%%
%%%%%%%%%%%%%%%%%%%%%%%%%%%%%%%%%%%%%%%%%%%%%%%%

The SQLS identifies lensed quasars among $\sim100,000$ quasars \citep{schneider10} 
from the SDSS Data Release 7 (DR7). The SDSS is a survey using a dedicated
wide-field 2.5-meter telescope \citep{gunn06} at the Apache Point
Observatory (New Mexico, USA) to conduct imaging and spectroscopic
surveys for a quarter of the sky. The imaging data were obtained
through five broad-band filters \citep[$ugriz$,][]{fukugita96,gunn98,doi10} 
with astrometric accuracy better than $\sim 0\farcs1$ \citep{pier03}
and with photometric zeropoint accuracy better than $\sim 0.01$~mag
\citep{hogg01,smith02,ivezic04,tucker06,padmanabhan08}. 
Quasar and galaxy candidates were selected by the SDSS spectroscopic
pipeline based on the imaging data
\citep{eisenstein01,richards02,strauss02}. 
The candidates were tiled in each plate according to the algorithm of
\citet{blanton03}, and were observed with multi-object fiber
spectrographs ($R\sim1800$) covering 3800{\,\AA} to 9200{\,\AA}
\citep{blanton03}. The SDSS DR7 data have already been made public 
\citep{stoughton02,abazajian03,abazajian04,abazajian05,abazajian09,
adelman06,adelman07,adelman08}. 

\begin{figure}
\epsscale{.45}
\plotone{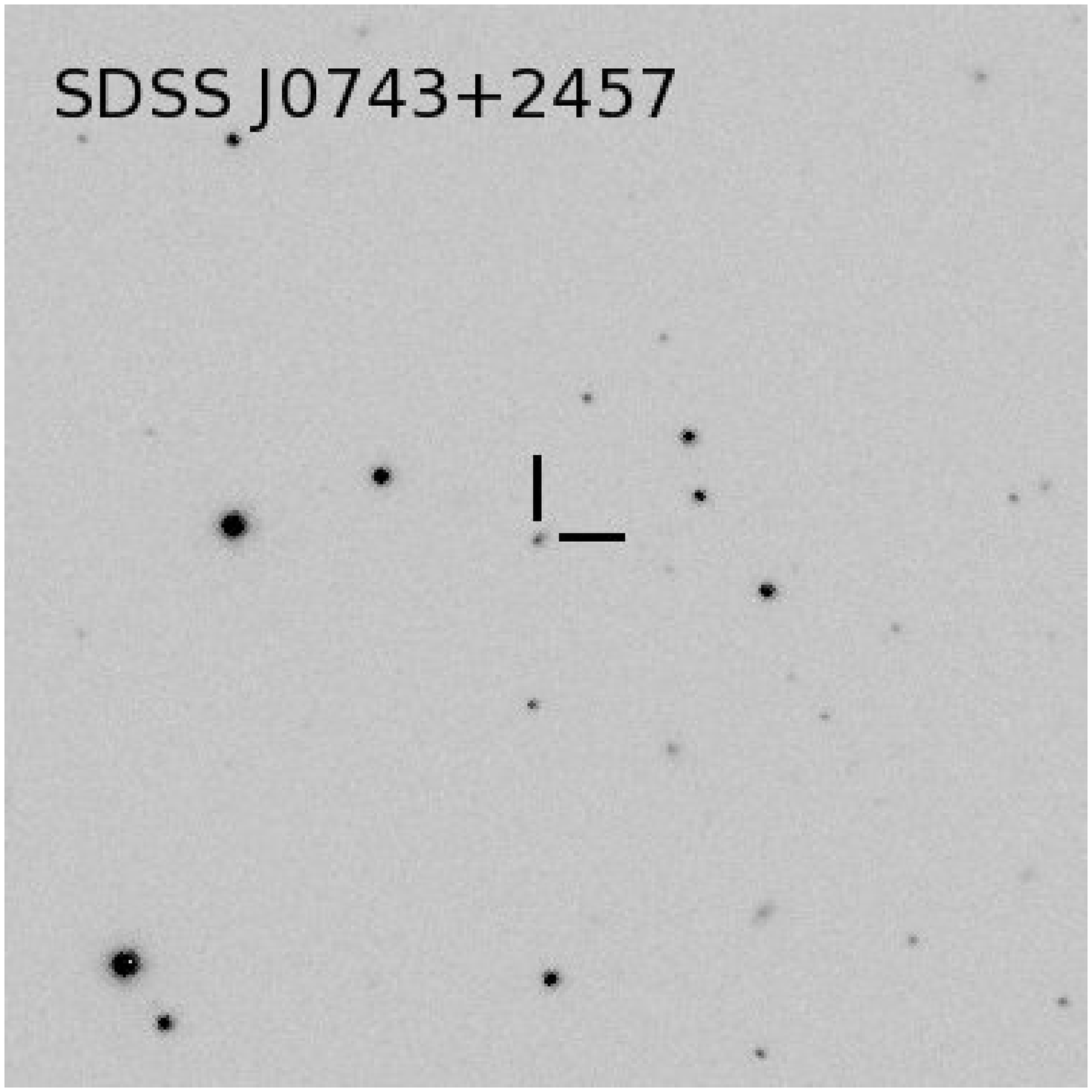}
\plotone{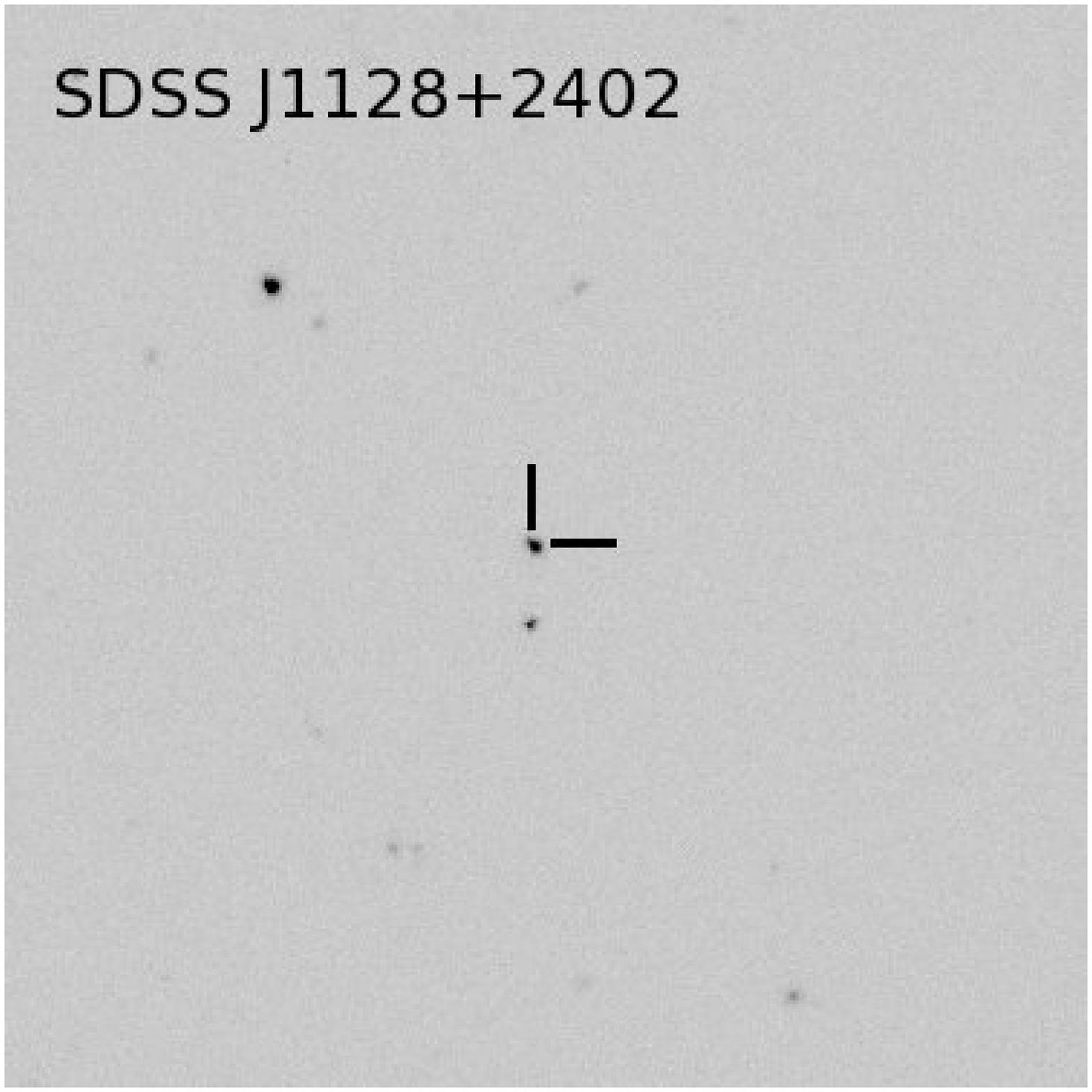}\\

\vspace*{0.2cm}

\plotone{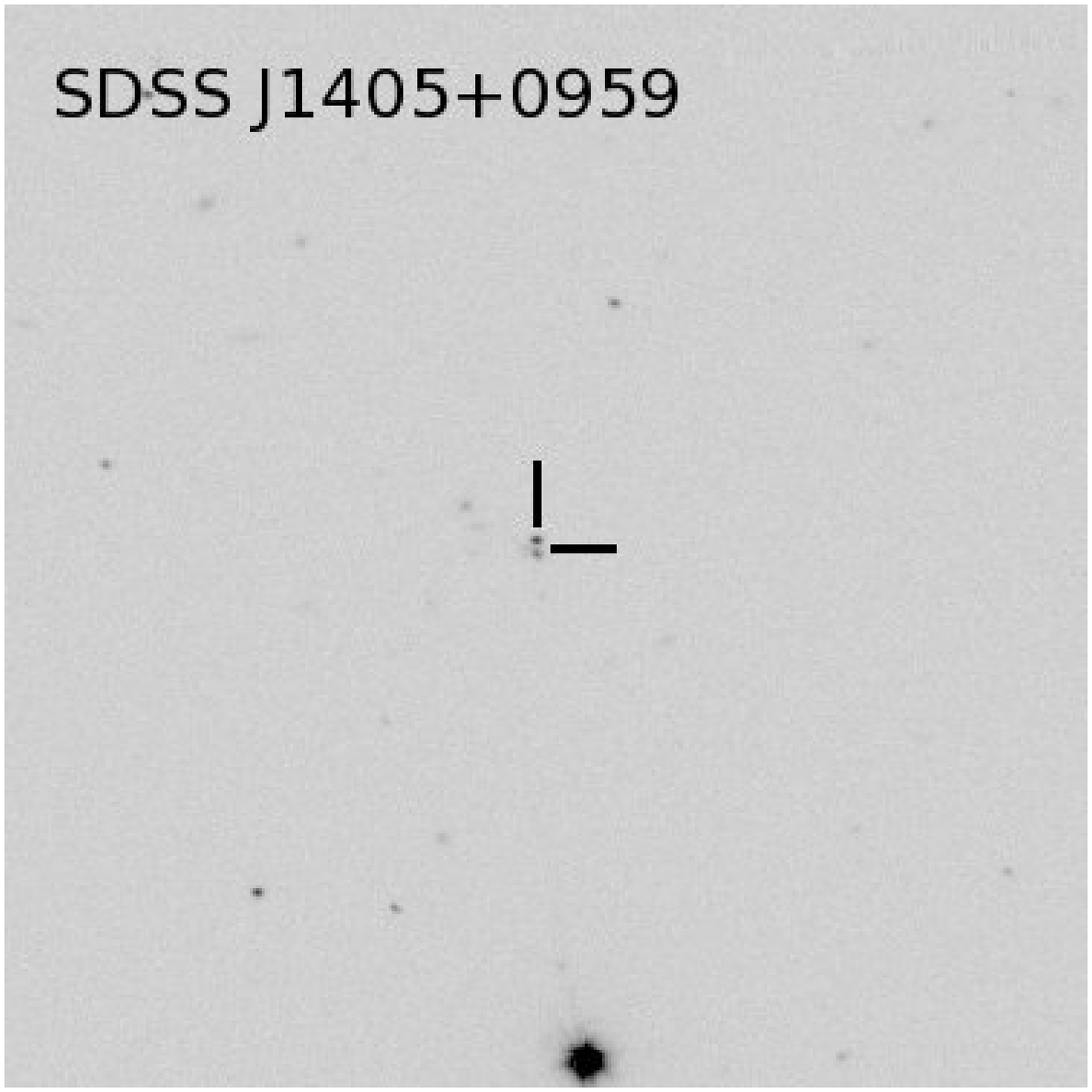}
\plotone{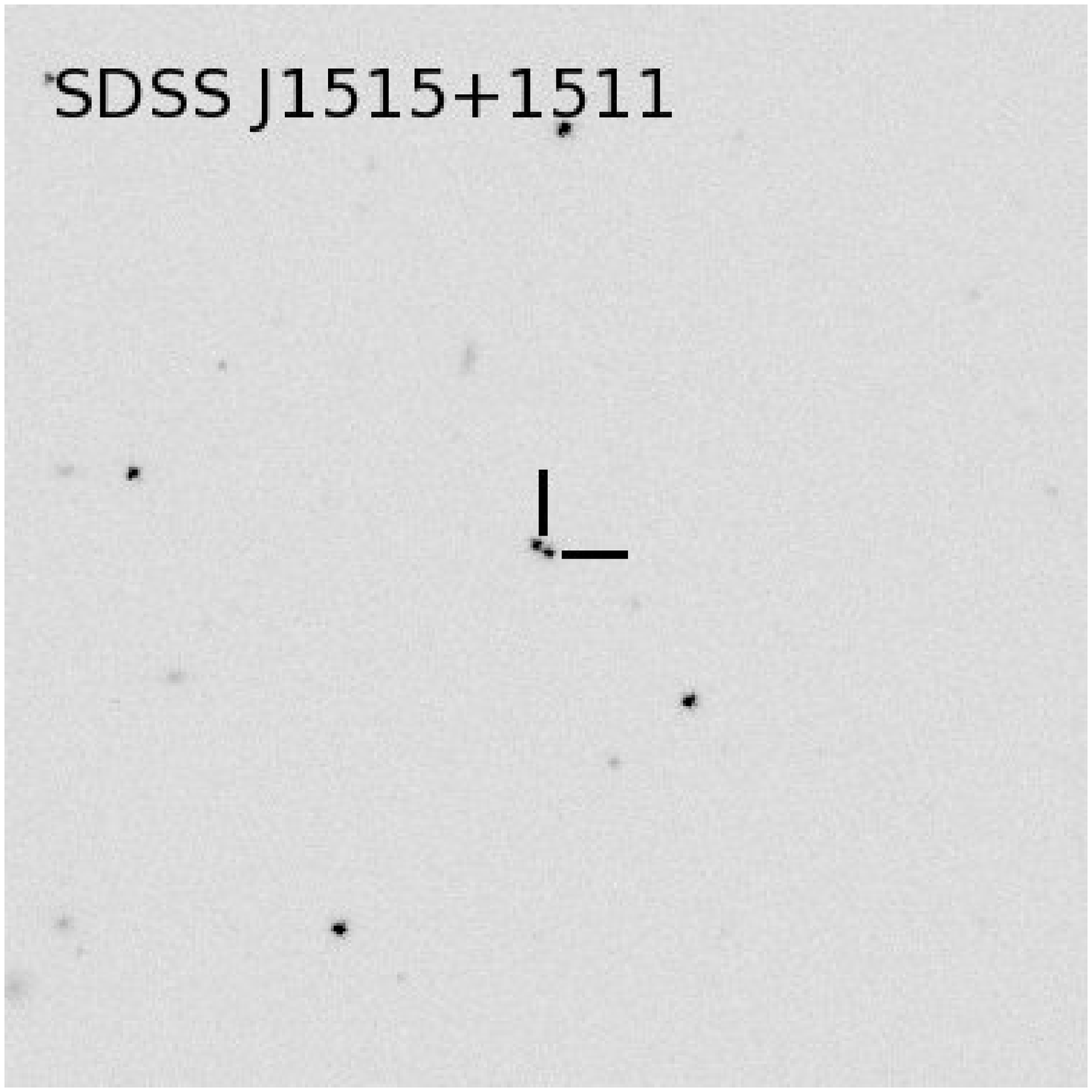}
\caption{Finding charts (SDSS $i$-band images) of the 4 lens systems. 
The size of each figure is $2\farcm5\times 2\farcm5$. North is up and 
East is left.
\label{fig:fc}}
\end{figure}

The procedure to identify lensed quasar candidates from the
spectroscopic SDSS quasars has been described in the series of SQLS
papers \citep{oguri06,inada08,inada10,inada12}, which we outline
here. We use two selection methods. One is to find quasars with nearby
objects whose colors are similar to the quasars. This method is tagged
as the ``color selection''. The color selection does not always work
well for the SDSS data because of the relatively poor spatial resolution  
of SDSS imaging data, whose seeing sizes are comparable to typical
image separations of quasar lenses. Therefore, in order to find lenses
which are not deblended into multiple components in the SDSS data, we
also exploit the so-called ``morphological selection'', in which we
select spectroscopic quasars with extended morphologies as lensed
quasar candidates. SDSS~J0743+2457 and SDSS~J1128+2402 are selected by
the morphological selection, whereas SDSS~J1405+0959 and
SDSS~J1515+1511 are selected by both the color selection and the
morphological selection. The SDSS properties of these four candidates
are summarized in Table~\ref{tab:sdss}. We also show the SDSS $i$-band
images of the four objects in Figure~\ref{fig:fc}.

We note that SDSS~J0743+2457 and SDSS~J1405+0959 have been reported as 
quasar lenses ULAS~074352.6+245743 and ULAS~140515.4+095931,
respectively, by the Major UKIDSS-SDSS Cosmic Lens Survey (MUSCLES) 
survey \citep{jackson12}. In the MUSCLES survey lens candidates are
selected by cross-matching SDSS quasars with UKIDSS near-infrared
images which typically have much better spatial resolution. Our
identification of SDSS~J0743+2457 and SDSS~J1405+0959 as lens
candidates is independent of the MUSCLES survey result. 

%%%%%%%%%%%%%%%%%%%%%%%%%%%%%%%%%%%%%%%%%%%%%%%%
%%%%%%%%%%%%%%%%%%%%%%%%%%%%%%%%%%%%%%%%%%%%%%%%
%%%%%%%%%%%%%%%%%%%%%%%%%%%%%%%%%%%%%%%%%%%%%%%%
\section{Follow-up Imaging and Spectroscopy}\label{sec:follow-up}
%%%%%%%%%%%%%%%%%%%%%%%%%%%%%%%%%%%%%%%%%%%%%%%%
%%%%%%%%%%%%%%%%%%%%%%%%%%%%%%%%%%%%%%%%%%%%%%%%
%%%%%%%%%%%%%%%%%%%%%%%%%%%%%%%%%%%%%%%%%%%%%%%%

%%%%%%%%%%%%%%%%%%%%%%%%%%%%%%%%%%%%%%%%%%%%%%%%
\subsection{Follow-up Data}
%%%%%%%%%%%%%%%%%%%%%%%%%%%%%%%%%%%%%%%%%%%%%%%%

We carried out follow-up imaging and spectroscopic observations using 
the University of Hawaii 2.2-meter telescope (UH88), the Astrophysical 
Research Consortium 3.5-meter telescope (ARC), the 3.58-meter Telescopio
Nazionale Galileo (TNG), the Subaru 8.2-m telescope, and the Gemini 
North 8.1-m telescope, to confirm the lensing hypothesis of the four
lensed quasar candidates. All the four candidates are doubly imaged
lens candidates, and therefore their lensing hypothesis should be
confirmed by both similar spectral energy distributions (SEDs) for the 
two stellar components and the existence of an extended object between
the two components that act as a lens. We summarize the details of our  
follow-up imaging and spectroscopic observations in Tables~\ref{tab:image} 
and \ref{tab:spec}. 

\begin{figure}
\epsscale{.45}
\plotone{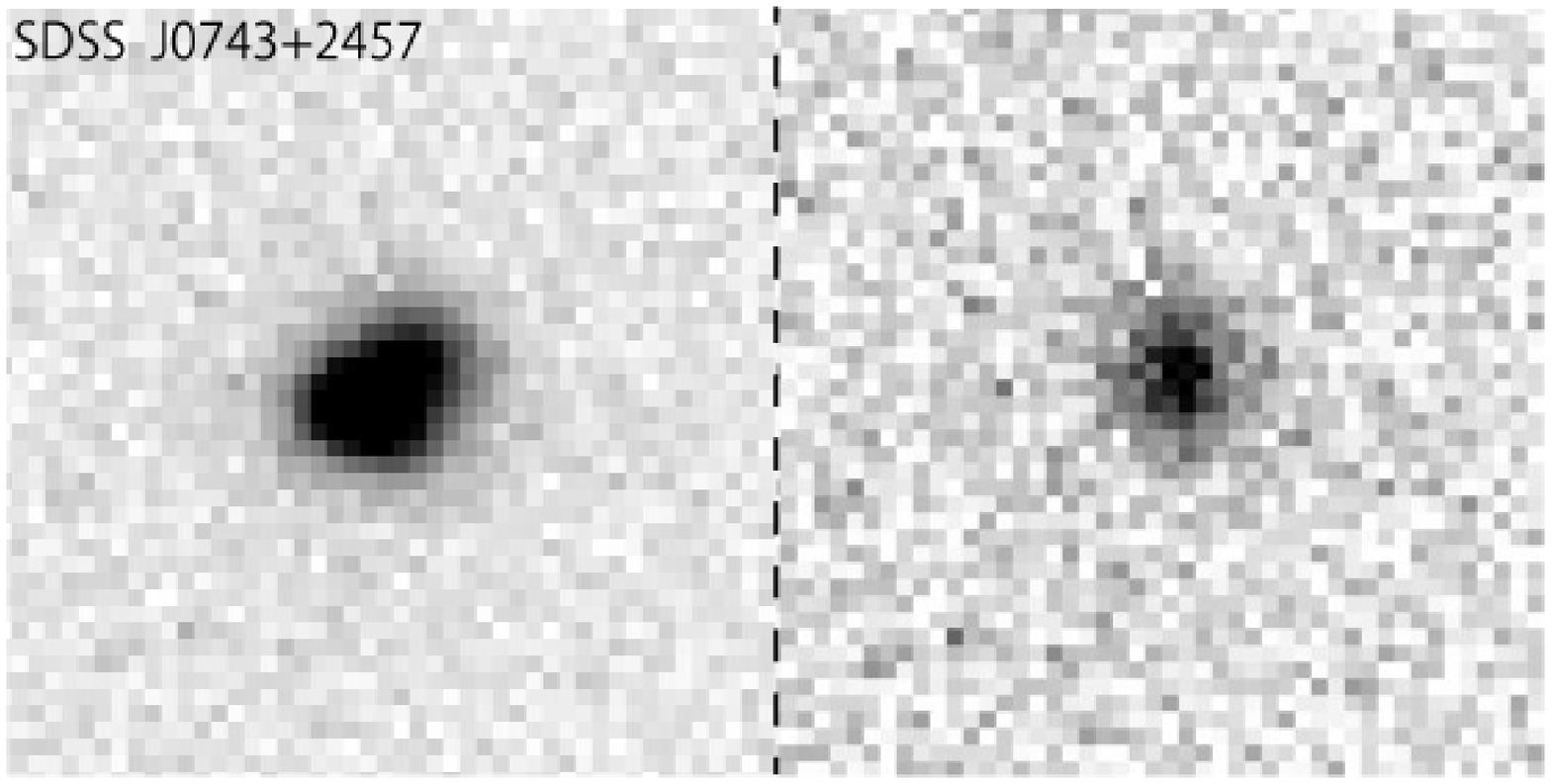}
\plotone{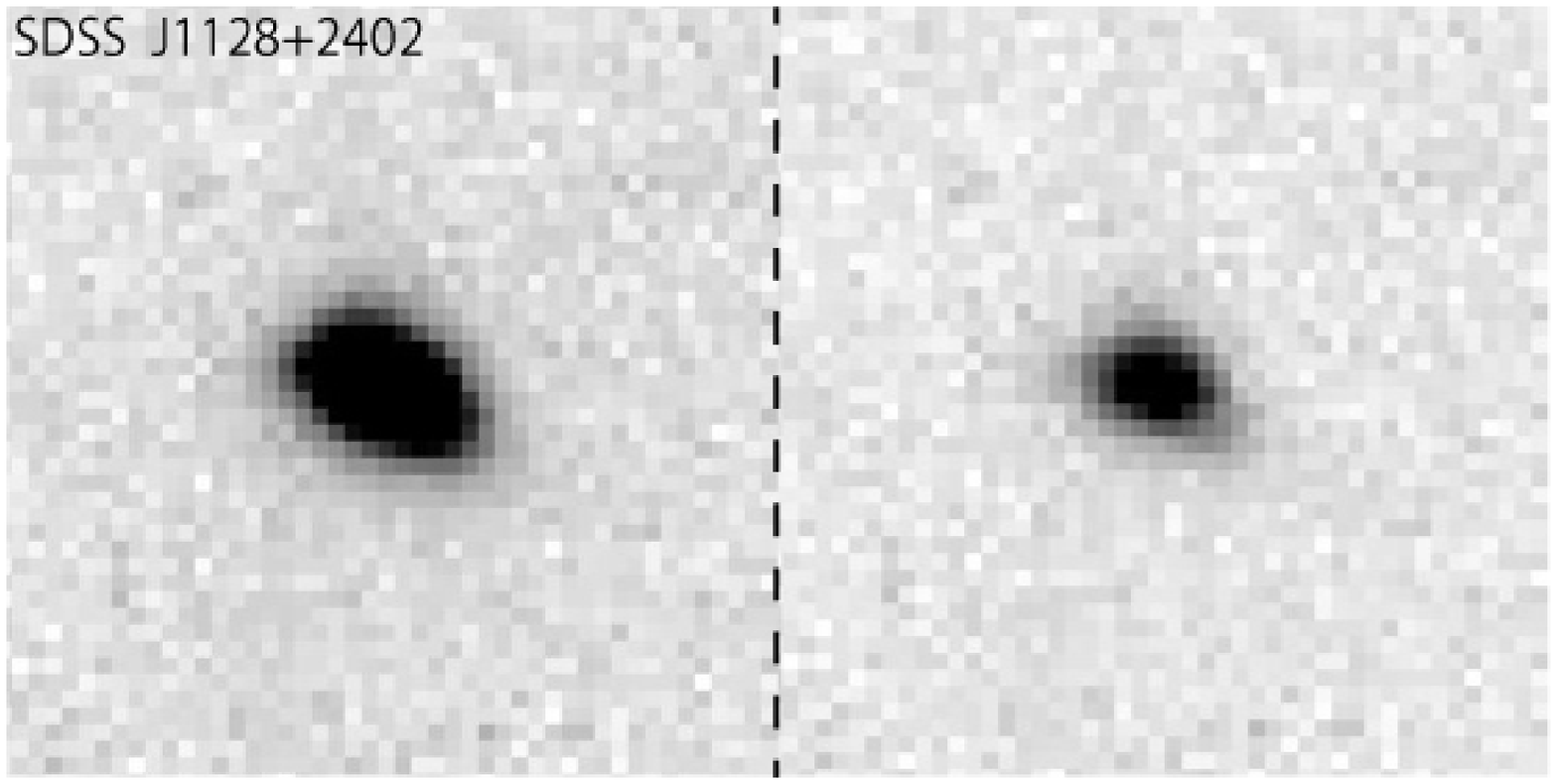}\\

\vspace*{0.2cm}

\plotone{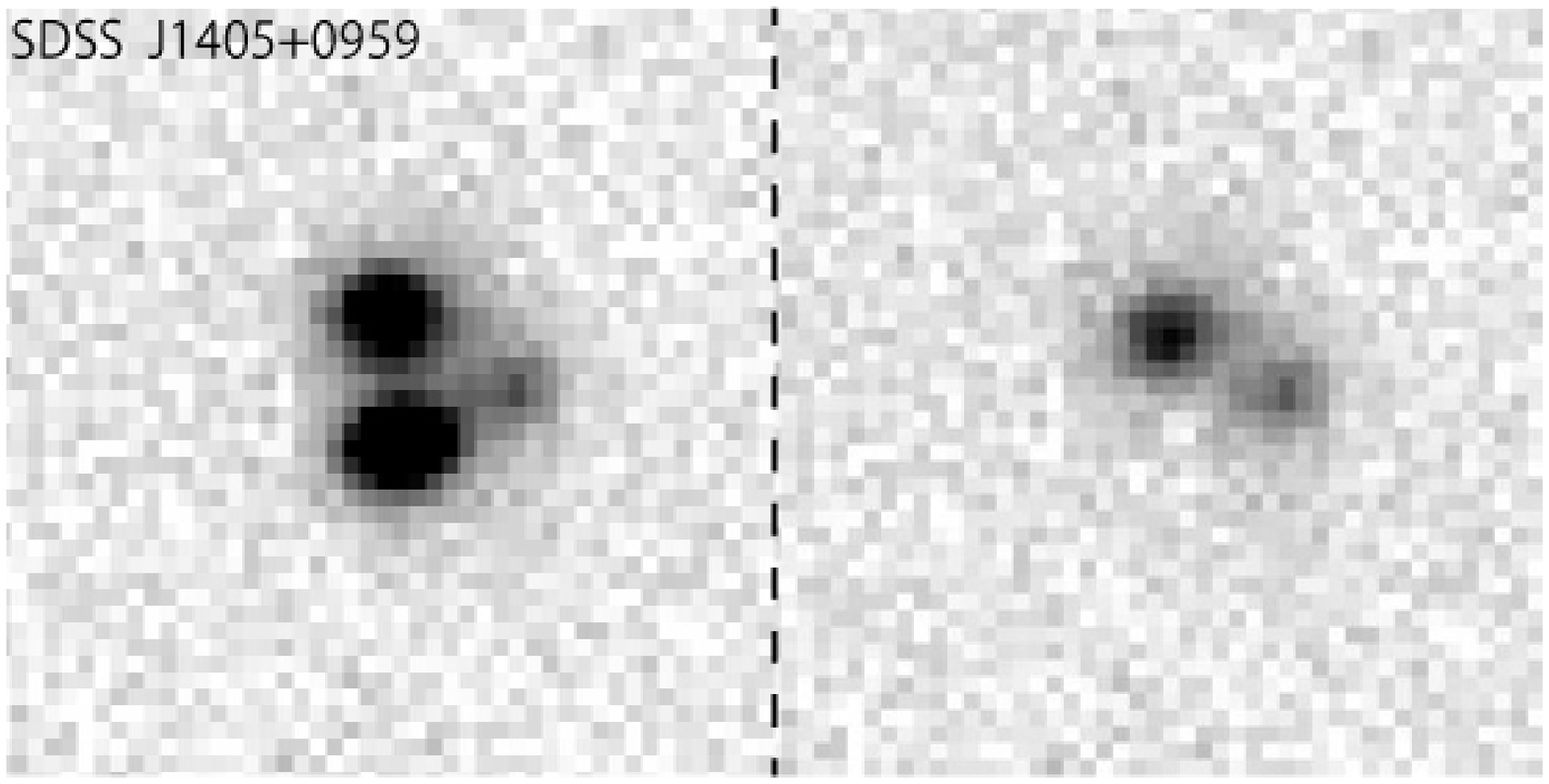}
\plotone{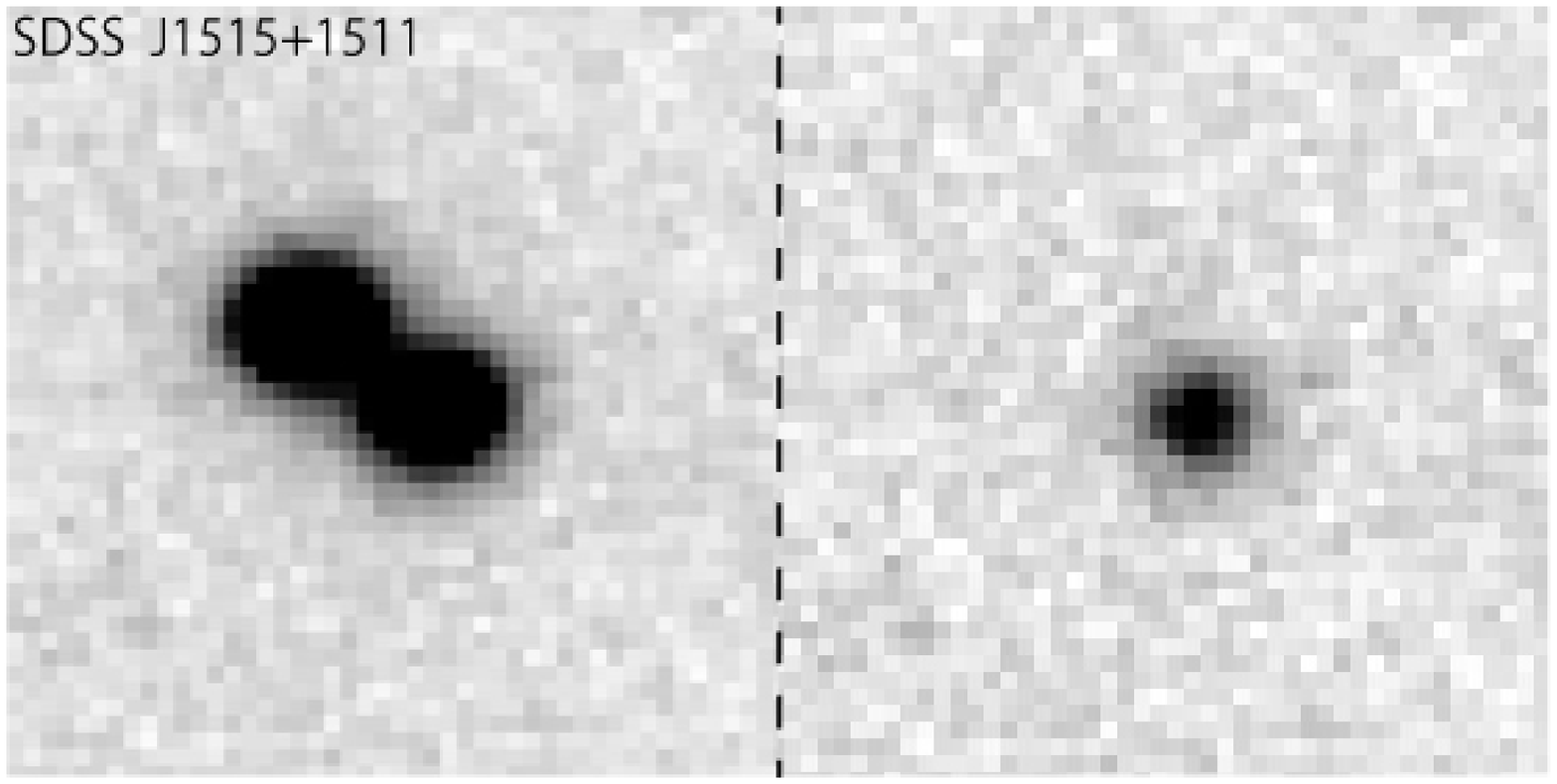}
\caption{
Follow-up $I$-band images of the 4 lens systems. The images of
SDSS~J0743+2457, SDSS~J1128+2402, and SDSS~J1515+1511 were taken with
UH88/Tek2k (0\farcs22 pixel${}^{-1}$), and the image of
SDSS~J1405+0959 were taken with UH88/8k (0\farcs235 pixel${}^{-1}$). 
North is up and East is left. For each system, the left panel shows
the original image and the right panel shows the image after
subtracting two PSF components. The residuals in the right panels are
all fitted well by the S\'{e}rsic profile. 
See Table~\ref{tab:sersic} for the fitting results.
\label{fig:iband}}
\end{figure}

We clearly detect two stellar components in each original image of the
follow-up imaging observations. Once we subtract two Point Spread
Functions (PSFs) using GALFIT \citep{peng02}, we detect a significant
residual, indicating an additional component in between
the two PSFs, for all four systems. We then model the system with two PSFs plus an extended
component represented by a S\'{e}rsic profile. As shown in
Figure~\ref{fig:iband}, an extended component is clearly visible after
subtracting the two PSFs. The S\'{e}rsic parameters of the
best-fitting model are summarized in Table~\ref{tab:sersic}.  We also
summarize the astrometric and photometric results for each system  in
Table~\ref{tab:opt}.  

\begin{figure}
\epsscale{.9}
\plotone{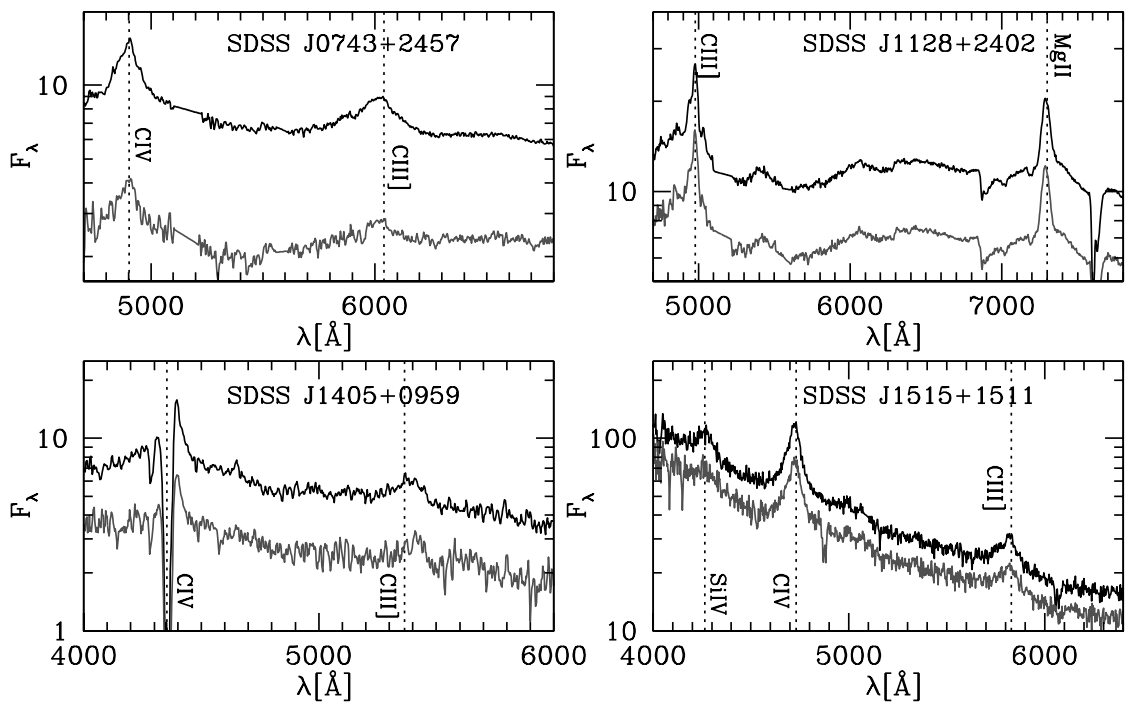}
\caption{
Spectra of two lensed quasar images. The data between 5100{\AA} and
5200{\AA} for SDSS~J0743+2457 and SDSS~J1128+2402 are excluded due to
a gap of the detector. The wavelengths of quasar emission lines are
indicated by vertical dotted lines. See Table~\ref{tab:spec} for
details of our  spectroscopic follow-up observations.
\label{fig:allspec}}
\end{figure}

Results of the spectroscopic follow-up observations are shown in
Figure~\ref{fig:allspec}. For each candidate, we aligned the slit
direction such that we can observe the two stellar components
simultaneously. The good seeing condition (FWHM$\lesssim$1\farcs0)
enabled us to extract the spectrum of each component easily using
standard IRAF\footnote{IRAF is distributed by the National Optical
  Astronomy Observatories, which are operated by the Association of
  Universities for Research in Astronomy, Inc., under cooperative
  agreement with the National Science Foundation.} tasks. The spectra
indicate that the SEDs of the two stellar components are quite
similar for all four systems. In particular, the two quasar
components of the SDSS~J1405+0959 system have very similar broad
absorption, which is in strong support of strong lensing
interpretation of this system. The fainter component of
SDSS~J1515+1511 shows a strong \ion{Mg}{2} absorption at $\sim
5900${\AA}, which corresponds to an absorber at $z=0.742$, and is
likely to be associated with the lensing galaxy (see below).

%%%%%%%%%%%%%%%%%%%%%%%%%%%%%%%%%%%%%%%%%%%%%%%%
\subsection{Lens Redshifts}
%%%%%%%%%%%%%%%%%%%%%%%%%%%%%%%%%%%%%%%%%%%%%%%%

\begin{figure}
\epsscale{.6}
\plotone{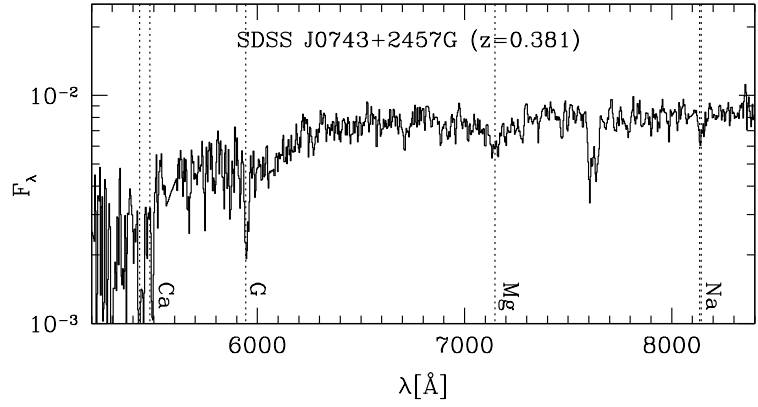}
\caption{The spectrum of the lensing galaxy G of
  SDSS~J0743+2457. The spectrum clearly indicates that the lensing
  galaxy is an early-type galaxy at $z=0.381$.
\label{fig:spec0743g}}
\end{figure}

First we examine our follow-up spectra carefully to look for any
spectral signature of lensing galaxies. We find that the spectrum
of the fainter component of SDSS~J0743+2457 contains a significant amount of 
galaxy flux. We extract the spectrum of the lensing
galaxy G by subtracting the spectrum of image A, with an appropriate
offset corresponding to the flux ratio, from that of image B. 
Figure~\ref{fig:spec0743g} clearly indicates that the lensing galaxy
is an early-type galaxy at $z_l=0.381$. 

For the other lens systems, we estimate redshifts of the lensing
galaxies based on their colors. For SDSS~J1405+0959, the S\'{e}rsic 
concentration index suggests that the residual may be a late-type
galaxy. In this case, the $V-R$ and $R-I$ colors imply that the
redshift is $z_l\sim 0.5$ \citep[e.g.,][]{fukugita95}. This is
broadly consistent with $z_l\sim 0.66$ derived in \citet{jackson12}
using the overall spectral shape of the lensing galaxy in their
follow-up spectroscopic data. For SDSS~J1515+1511, the morphology is
more robust in the $i$-band, where it is consistent with a late-type galaxy.
Its color $i-K'\sim4.4$ implies the redshift of $z\sim 0.8-1.1$, which
is close to the redshift $z=0.742$ of the strong absorber seen in the
spectrum of the fainter quasar image. Thus we tentatively assign its
lens redshift to be $z_l=0.742$.  It is difficult to estimate the
photometric lens redshift of SDSS~J1128+2402 from the current data.

%%%%%%%%%%%%%%%%%%%%%%%%%%%%%%%%%%%%%%%%%%%%%%%%
%%%%%%%%%%%%%%%%%%%%%%%%%%%%%%%%%%%%%%%%%%%%%%%%
%%%%%%%%%%%%%%%%%%%%%%%%%%%%%%%%%%%%%%%%%%%%%%%%
\section{Mass Modeling}\label{sec:model}
%%%%%%%%%%%%%%%%%%%%%%%%%%%%%%%%%%%%%%%%%%%%%%%%
%%%%%%%%%%%%%%%%%%%%%%%%%%%%%%%%%%%%%%%%%%%%%%%%
%%%%%%%%%%%%%%%%%%%%%%%%%%%%%%%%%%%%%%%%%%%%%%%%

We model the lens systems adopting a Singular Isothermal Ellipsoid
(SIE) mass model using {\it glafic} \citep{oguri10}. The number of
parameters is eight (the position and flux of the source quasar, the
position of the lensing galaxy, the Einstein radius, the ellipticity,
and the position angle), while the number of constraints from the
imaging observations is  also eight (the positions and fluxes of the two
quasar components, and the position of the lensing galaxy). Thus the
model has zero degrees of freedom. We adopt the $I$-band imaging result 
as constraints for SDSS~J0743+2457, SDSS~J1128+2402, and SDSS~J1405+0959. 
For SDSS~J1515+1511, we use the $i$-band imaging result in spite of the lens being 
comparatively fainter, because the seeing was smaller, there were more stars 
available to build a PSF, and both astrometry and morphology are more robust.
 We summarize the parameters of the best-fitting
($\chi^2\sim 0$) models in Table~\ref{tab:model}. As is common
\citep[e.g.,][]{keeton98}, there are large differences in the lensing
galaxy shapes between the observations (Table~\ref{tab:sersic}) and the 
best-fitting mass model parameters (Table~\ref{tab:model}). This may partly be
explained by nearby perturbers which affect the lens potentials.

%%%%%%%%%%%%%%%%%%%%%%%%%%%%%%%%%%%%%%%%%%%%%%%%
%%%%%%%%%%%%%%%%%%%%%%%%%%%%%%%%%%%%%%%%%%%%%%%%
%%%%%%%%%%%%%%%%%%%%%%%%%%%%%%%%%%%%%%%%%%%%%%%%
\section{Summary}\label{sec:summary}
%%%%%%%%%%%%%%%%%%%%%%%%%%%%%%%%%%%%%%%%%%%%%%%%
%%%%%%%%%%%%%%%%%%%%%%%%%%%%%%%%%%%%%%%%%%%%%%%%
%%%%%%%%%%%%%%%%%%%%%%%%%%%%%%%%%%%%%%%%%%%%%%%%
   
We have confirmed the lensing hypothesis for four lens candidates
selected by the SQLS, based on the similar SEDs of the stellar components
together with the existence of extended objects in between the stellar
components, and the successful mass modeling of the observed configurations. 
All four are doubly imaged quasar lens systems:
SDSS~J0743+2457 ($z_s=2.165$, $z_l=0.381$, $\theta=1\farcs034$),
SDSS~J1128+2402 ($z_s=1.608$, $\theta=0\farcs844$),  
SDSS~J1405+0959 ($z_s=1.810$, $z_l \sim 0.66$, $\theta=1\farcs978$), 
and SDSS~J1515+1511 ($z_s=2.054$, $z_l =0.742$, $\theta=1\farcs989$). 
Two of them (SDSS~J1405+0959 and SDSS~J1515+1511) are included in the
SQLS statistical sample \citep{inada12} and are used for cosmological
constraints \citep{oguri12}. 

These lens systems represent likely the last quasar lenses to emerge from the
SQLS project. The applications of the SQLS lenses have been limited by
the lack of high-resolution images for many of the SQLS lenses. We are therefore
currently conducting a program to observe the SQLS lenses with Subaru Telescope
laser guide start adaptive optics system to accurately measure the
quasar image positions and light profile of the lensing galaxy
\citep[][C. E. Rusu et al., in preparation]{rusu11}. We are also
carrying out a lensed quasar survey using the SDSS-III data
\citep{eisenstein11} from which we will be able to discover more
quasar lenses. 

\acknowledgments

This work was supported in part by the FIRST program
``Subaru Measurements of Images and Redshifts (SuMIRe)'', World
Premier International Research Center Initiative (WPI Initiative),
MEXT, Japan, and Grant-in-Aid for Scientific Research from the JSPS 
(23740161). I.~K. acknowledges support from MEXT KAKENHI 24740171.
Use of the UH 2.2-m telescope for the observations
is supported by NAOJ. Based in part on observations obtained with the Apache 
Point Observatory 3.5-meter telescope, which is owned and operated by 
the Astrophysical Research Consortium, and on observations made with the Italian 
Telescopio Nazionale Galileo (TNG) operated on the island of La Palma by the 
Fundacion Galileo Galilei of the INAF (Istituto Nazionale di Astrofisica) at 
the Spanish Observatorio del Roque de los Muchachos of the Instituto de 
Astrofisica de Canarias. Based in part on observations obtained at the Gemini 
Observatory, which is operated by the Association of Universities for Research 
in Astronomy, Inc., under a cooperative agreement with the NSF on behalf of the 
Gemini partnership: the National Science Foundation (United States), 
the National Research Council (Canada), CONICYT (Chile), the Australian Research  
Council (Australia), Minist\'{e}rio da Ci\^{e}ncia, Tecnologia e Inova\c{c}\~{a}o (Brazil) and
Ministerio de Ciencia, Tecnolog\'{i}a e Innovaci\'{o}n Productiva (Argentina).

Funding for the SDSS and SDSS-II has been provided by the Alfred
P. Sloan Foundation, the Participating Institutions, the National
Science Foundation, the U.S. Department of Energy, the National
Aeronautics and Space Administration, the Japanese Monbukagakusho, the
Max Planck Society, and the Higher Education  Funding Council for
England. The SDSS Web Site is http://www.sdss.org/. 

The SDSS is managed by the Astrophysical Research Consortium for the
Participating Institutions. The Participating Institutions are the
American Museum of Natural History, Astrophysical Institute Potsdam,
University of Basel, Cambridge University, Case Western Reserve
University, University of Chicago, Drexel University, Fermilab, the
Institute for Advanced Study, the Japan Participation Group, Johns
Hopkins University, the Joint Institute for Nuclear Astrophysics, the
Kavli Institute for Particle Astrophysics and Cosmology, the Korean
Scientist Group, the Chinese Academy of Sciences (LAMOST), Los Alamos
National Laboratory, the Max-Planck-Institute for Astronomy (MPIA),
the Max-Planck-Institute for Astrophysics (MPA), New Mexico State
University, Ohio State University, University of Pittsburgh,
University of Portsmouth, Princeton University, the United States
Naval Observatory, and the University of Washington.

\clearpage

%%%%%%%%%%%%%%%%%%%%%%%%%%%%%%%%%%%%%%%%%%%%%%%%
\begin{deluxetable}{ccccccccc}
\tabletypesize{\scriptsize}
\rotate
\tablecaption{SDSS Properties of the Lens Systems\label{tab:sdss}}
\tablewidth{0pt}
\tablehead{ \colhead{Object} & \colhead{R.A. (J2000)} &
\colhead{Decl. (J2000)} & \colhead{$u$} & \colhead{$g$}
& \colhead{$r$} & \colhead{$i$} & \colhead{$z$} & \colhead{Redshift}} 
\startdata
SDSS J0743+2457 & 07:43:52.61 & +24:57:43.6 & $19.92\pm0.04$ & $19.42\pm0.02$ &
 $19.19\pm0.02$ & $19.09\pm0.02$ & $18.76\pm0.04$ & $2.165$ \\ \hline
SDSS J1128+2402 & 11:28:18.49 & +24:02:17.4 & $18.35\pm0.02$ & $18.30\pm0.02$ &
 $18.12\pm0.02$ & $17.92\pm0.01$ & $17.92\pm0.02$ & $1.608$ \\ \hline
SDSS J1405+0959 & 14:05:15.42 & +09:59:31.3 & $19.97\pm0.06$ & $19.62\pm0.04$ &
 $19.48\pm0.03$ & $19.10\pm0.04$ & $19.19\pm0.06$ & $1.810$\\ 
                & 14:05:15.40 & +09:59:29.3 & $20.67\pm0.15$ & $20.52\pm0.13$ &
 $20.31\pm0.09$ & $19.72\pm0.09$ & $19.50\pm0.12$ & \nodata \\ \hline
SDSS J1515+1511 & 15:15:38.59 & +15:11:35.8 & $18.43\pm0.04$ & $18.27\pm0.04$ &
 $18.23\pm0.04$ & $18.11\pm0.04$ & $17.91\pm0.04$ & $2.054$ \\ 
                & 15:15:38.47 & +15:11:34.7 & $18.94\pm0.09$ & $18.70\pm0.08$ &
 $18.61\pm0.07$ & $18.37\pm0.07$ & $18.19\pm0.07$ & \nodata \\
\enddata
\end{deluxetable}
%%%%%%%%%%%%%%%%%%%%%%%%%%%%%%%%%%%%%%%%%%%%%%%%

\clearpage

%%%%%%%%%%%%%%%%%%%%%%%%%%%%%%%%%%%%%%%%%%%%%%%%
\begin{deluxetable}{cccc}
\tabletypesize{\scriptsize}
\rotate
\tablecaption{Summary of Imaging Follow-up Observations\label{tab:image}}
\tablewidth{0pt}
\tablehead{ \colhead{Object} & \colhead{Instruments} &
\colhead{Observing Date (UT)} & \colhead{Exposure (s)}}
\startdata
SDSS J0743+2457 & Tek2k($VRIz$)  & 2010Jan16($V$), 
2009Apr16($R$), 2010Jan17($Iz$) & 720($V$), 300($R$), 960($Iz$)\\
SDSS J1128+2402 & Tek2k($VI$)   & 2007Apr12($V$), 2010Jan17($I$)
& 300($V$), 960($I$)\\
SDSS J1405+0959 & UH8k($VI$), Tek2k($R$) & 2006Mar25($VI$), 
2009Apr16($R$) & 270($V$), 400($R$), 450($I$)\\
SDSS J1515+1511 & Tek2k($I$), S-cam($i$), IRCS($K'$), SPIcam($z$)
& 2009Apr16($I$), 2011Mar31($i$), 2011May7($K'$), 2011Mar26($z$) & 400($I$), 90($i$), 1320($K'$), 1080($z$) \\
\enddata
\tablecomments{The Tektronix 2048$\times$2048 CCD camera (Tek2k) 
and the UH8k wide-field imager (UH8k) are installed on the UH88 telescope. 
The Suprime-Cam \citep[S-cam;][]{miyazaki02} and 
the Infrared Camera and Spectrograph \citep[IRCS;][]{kobayashi00} are installed on 
the Subaru telescope. The Seaver Prototype Imaging camera (SPIcam) is installed on 
the ARC telescope.}
\end{deluxetable}
%%%%%%%%%%%%%%%%%%%%%%%%%%%%%%%%%%%%%%%%%%%%%%%%

\clearpage

%%%%%%%%%%%%%%%%%%%%%%%%%%%%%%%%%%%%%%%%%%%%%%%%
\begin{deluxetable}{ccccccc}
\tabletypesize{\scriptsize}
\rotate
\tablecaption{Summary of Spectroscopic Follow-up Observations\label{tab:spec}}
\tablewidth{0pt}
\tablehead{ \colhead{Object} & \colhead{Instrument} &
\colhead{Slit width} & \colhead{Filter/Grism} & \colhead{Wavelength Coverage
(\AA)} & \colhead{Observing Date (UT)} & \colhead{Exposure (s)} }
\startdata
SDSS J0743+2457 & GMOS  & $1''$ & GG455/R150 & 4600-9000 & 2010 Mar 22 & 1200 \\
SDSS J1128+2402 & GMOS  & $1''$ & GG455/R150 & 4600-9000 & 2010 Mar 22 & 900 \\
SDSS J1405+0959 & DOLORES  & $1''$ & LR-B & 3700-7800 & 2008 Mar 1 & 1800\\
SDSS J1515+1511 & DOLORES  & $1''$ & LR-B & 3700-7800 & 2008 Apr 13 & 900\\
\enddata
\tablecomments{The Gemini Multi-Object Spectrograph (GMOS)
is installed on the Gemini telescope (North). The Device 
Optimized for the LOw RESolution (DOLORES) is installed on the TNG telescope. }
\end{deluxetable}
%%%%%%%%%%%%%%%%%%%%%%%%%%%%%%%%%%%%%%%%%%%%%%%%

\clearpage

%%%%%%%%%%%%%%%%%%%%%%%%%%%%%%%%%%%%%%%%%%%%%%%%
\begin{deluxetable}{ccccc}
\tablewidth{0pt}
\tablecaption{S\'{e}rsic Parameters of the Lensing Galaxies\label{tab:sersic}}
\tablehead{\colhead{Object} & \colhead{$r_e$\tablenotemark{a}(${}''$)} &
 \colhead{$n$\tablenotemark{b}} & \colhead{$e$\tablenotemark{c}} 
 & \colhead{{$\theta_e{({}^{\circ})}$}\tablenotemark{c} }
}
\startdata
SDSS~J0743+2457  &  0.58{$\pm$}0.24  &  0.71{$\pm$}0.85  &  0.47{$\pm$}0.16  &  6.48{$\pm$}14.65  \\
SDSS~J1128+2402  &  0.33{$\pm$}0.06  &  1.79{$\pm$}0.50  &  0.49{$\pm$}0.03  &  56.03{$\pm$}3.56  \\
SDSS~J1405+0959  &  0.37{$\pm$}0.05  &  2.15{$\pm$}0.77  &  0.06{$\pm$}0.11  &  $-$17.02{$\pm$}81.33  \\
SDSS~J1515+1511  &  1.06{$\pm$}0.08  &  1.08{$\pm$}0.20  &  0.53{$\pm$}0.03  &  $-16.20${$\pm$}2.50  \\
\enddata
\tablecomments{Best-fitting S\'{e}rsic parameters measured in the $I$-band 
images with GALFIT (The parameters for SDSS~J1515+1511 are measured in the $i$-band. 
The Subaru S-cam $i$-band image has smaller seeing).}
\tablenotetext{a}{Effective radius (in arcsec) of the S\'{e}rsic profile.}
\tablenotetext{b}{S\'{e}rsic concentration index.}
\tablenotetext{c}{Ellipticity and position angle measured East of North.}
\end{deluxetable}
%%%%%%%%%%%%%%%%%%%%%%%%%%%%%%%%%%%%%%%%%%%%%%%%

\clearpage

%%%%%%%%%%%%%%%%%%%%%%%%%%%%%%%%%%%%%%%%%%%%%%%%
\begin{deluxetable}{cccccc}
\tabletypesize{\scriptsize}
%\rotate
\tablecaption{Relative Astrometry and Photometry of the Lens Systems\label{tab:opt}}
\tablewidth{0pt}
\tablehead{ \colhead{Object} & \colhead{$\Delta X$ [arcsec]} &
\colhead{$\Delta Y$ [arcsec]} & \colhead{$V$} & \colhead{$R$} & 
\colhead{$I$} }
\startdata
\multicolumn{6}{c}{SDSS J0743+2457 ($\theta=1\farcs034\pm 0.056$)} \vspace*{1.5mm} \\ \hline \vspace*{-2.0mm} \\ 
A &  $\equiv 0.000$ & $\equiv0.000$   &  $19.33 \pm 0.01$ & $19.14 \pm
0.01$  & $18.90 \pm 0.02$ \\
B & $-0.796 \pm 0.040$ & $-0.660 \pm 0.040$ & $21.03 \pm 0.06$ &
$20.53 \pm 0.02$ &$20.36 \pm 0.31$  \\
G & $-0.590 \pm 0.070$ & $-0.546 \pm 0.070$  & \nodata & \nodata &
$20.47 \pm 0.37$ \\
\cutinhead{SDSS J1128+2402 ($\theta=0\farcs844\pm0\farcs006$)} 
A &  $\equiv0.000$    & $\equiv0.000$    & $18.55 \pm 0.01$ & \nodata
& $18.48 \pm 0.02$ \\
B & $-0.669 \pm 0.004$& $0.515 \pm 0.004$& $19.38 \pm 0.01$ & \nodata
& $19.15 \pm 0.03$  \\ 
G & $-0.429 \pm 0.007$& $0.361 \pm 0.007$& $21.74 \pm 0.25$ & \nodata
& $18.56 \pm 0.06$  \\ 
\cutinhead{SDSS J1405+0959 ($\theta=1\farcs978\pm0\farcs013$)}
A &   $\equiv0.000 $ &  $\equiv0.000$         & $19.34 \pm 0.01$ &
$19.18 \pm 0.01$ &  $18.71 \pm 0.01$ \\
B  &  $0.282 \pm 0.009$ &  $-1.958 \pm 0.009$ & $20.43 \pm 0.01$ &
$20.22 \pm 0.04$ &  $19.90 \pm 0.05$ \\ 
G1 &  $0.021 \pm 0.021$ &  $-1.614 \pm 0.021$ & $22.42 \pm 0.20$ &
$20.59 \pm 0.04$ &  $19.70 \pm 0.05$ \\ 
G2&   $-1.553 \pm 0.024$ &  $-0.792 \pm 0.024$&  $22.51 \pm 0.67$ &
$21.25 \pm 0.09$ &  $20.53 \pm 0.10$ \\
\cutinhead{SDSS J1515+1511 ($\theta=1\farcs989\pm0\farcs0.003$)}
A &   $\equiv0.000 $ &  $\equiv0.000$         & 
$i=18.15 \pm 0.01$ & $z=17.89 \pm 0.01$ & $K'=15.80 \pm 0.01$ \\
B  &  $1.648 \pm 0.002$ &  $-1.114 \pm 0.002$ &
$i=18.54 \pm 0.01$ & $z=18.32 \pm 0.02$ & $K'=16.06 \pm 0.01$ \\ 
G &  $1.384 \pm 0.030$ &  $-0.946 \pm 0.036$ & 
$i=21.39 \pm 0.06$ & $z=20.73 \pm 0.51$ & $K'=17.03 \pm 0.04$ \\ 
\enddata
\tablecomments{Magnitudes are in the Vega system, except for $i$- and $z$-band 
  which are in AB system. The $K'$ band photometric zero point for SDSS~J1515+1511 is 
  less robust, as it is based on a single star in the field of view (non-photometric 
  conditions), with catalog uncertainty 0.11~mag.}
\end{deluxetable}
%%%%%%%%%%%%%%%%%%%%%%%%%%%%%%%%%%%%%%%%%%%%%%%%

\clearpage

%%%%%%%%%%%%%%%%%%%%%%%%%%%%%%%%%%%%%%%%%%%%%%%%
\begin{deluxetable}{ccccc}
\tablewidth{0pt}
\tablecaption{Results of Mass Modeling\label{tab:model}}
\tablehead{\colhead{Object} & \colhead{$R_{\rm E}$(${}''$)\tablenotemark{a}} &
 \colhead{$e$\tablenotemark{b}} &
 \colhead{$\theta_e{({}^{\circ})}$\tablenotemark{b}} & \colhead{{$\mu_{\rm tot}$\tablenotemark{c}} }
} 
\startdata
SDSS~J0743+2457  &  0.53  &  0.19  &  89.4     &  3.6  \\
SDSS~J1128+2402  &  0.42  &  0.06  & $-$75.4   &  6.0  \\
SDSS~J1405+0959  &  1.03  &  0.49  &  67.4     &  2.6  \\
SDSS~J1515+1511  &  0.90  &  0.60  & $-$25.0   &  2.8  \\
\enddata
\tablenotetext{a}{Einstein radius in arcsec.}
\tablenotetext{b}{Ellipticity and position angle measured East of North.}
\tablenotetext{c}{Total magnification factor predicted by the best
  fitting model.}
\end{deluxetable}
%%%%%%%%%%%%%%%%%%%%%%%%%%%%%%%%%%%%%%%%%%%%%%%%

\clearpage

%%%%%%%%%%%%%%%%%%%%%%%%%%%%%%%%%%%%%%%%%%%%%%%%%%%%%%%%%%%%%%%%%%%%%%%
%%%%%%%%%%%%%%%%%%%%%%%%%%%%%%%%%%%%%%%%%%%%%%%%%%%%%%%%%%%%%%%%%%%%%%%

\clearpage

%%%%%%%%%%%%%%%%%%%%%%%%%%%%%%%%%%%%%%%%%%%%%%%%%%%%%%%%%%%%%%%%%%%%%%%
%%%%%%%%%%%%%%%%%%%%%%%%%%%%%%%%%%%%%%%%%%%%%%%%%%%%%%%%%%%%%%%%%%%%%%%

\clearpage

%%%%%%%%%%%%%%%%%%%%%%%%%%%%%%%%%%%%%%%%%%%%%%%%%%%%%%%%%%%%%%%%%%%%%%%
%%%%%%%%%%%%%%%%%%%%%%%%%%%%%%%%%%%%%%%%%%%%%%%%%%%%%%%%%%%%%%%%%%%%%%%

\clearpage

%%%%%%%%%%%%%%%%%%%%%%%%%%%%%%%%%%%%%%%%%%%%%%%%%%%%%%%%%%%%%%%%%%%%%%%
%%%%%%%%%%%%%%%%%%%%%%%%%%%%%%%%%%%%%%%%%%%%%%%%%%%%%%%%%%%%%%%%%%%%%%%

\end{document}